\documentclass[preprint,12pt]{elsarticle}
\usepackage{eurosym}
\usepackage{float}
\usepackage{amsfonts}
\usepackage{mdframed}
\usepackage{amssymb}
\usepackage{amsmath}
\usepackage{graphicx}
\usepackage[font={footnotesize,it}]{caption}

\begin{document}

\date{\today}

\begin{frontmatter}
\title{Absorption Cross-section and Decay Rate of Rotating Linear Dilaton
Black Holes}
\author{I. Sakalli \corref{cor1}}
\ead{izzet.sakalli@emu.edu.tr}
\author{O. A. Aslan  \corref{cor2}}
\ead{onur.aslan@emu.edu.tr}
\address{Physics Department , Eastern Mediterranean University,
Famagusta, Northern Cyprus, Mersin 10, Turkey.}

\begin{abstract}
We analytically study the scalar perturbation of non-asymptotically flat
(NAF) rotating linear dilaton black holes (RLDBHs) in 4-dimensions. We show
that both radial and angular wave equations can be solved in terms of the
hypergeometric functions. The exact greybody factor (GF), the absorption
cross-section (ACS), and the decay rate (DR) for the massless scalar waves are
computed for these black holes (BHs). The results obtained for ACS and DR
are discussed through graphs.

\end{abstract}
\end{frontmatter}

\section{Introduction}

Hawking's semiclassical study \cite{HR1} on BHs showed that BHs emit
particles from their "edge", known as the event horizon. This phenomenon is
known as Hawking radiation (HR), named after Hawking. In fact, HR arises
from the steady conversion of quantum vacuum fluctuations around into the pairs
of particles, one of which escaping at spatial infinity (SI) while the other
is trapped inside the event horizon. Calculations of HR reveal a
characteristic blackbody spectrum. Thus, putting quantum mechanics and
general relativity into the process, BHs become no longer "black" but obey
the laws of thermodynamics. However, the spacetime geometry around BH
modifies HR by the so-called GFs. Namely, an observer at SI detects not a perfect black body spectrum but a modification of
this since GFs are dependent upon both geometry and frequency \cite%
{Maldacena}.

The first papers of GFs (and its related subjects: ACS and DR) date back
to the nineteen-seventies \cite{Star, Chur,Page1,Page2,Unruh}. Today,
although there exists numerous studies on the subject (see for example \cite%
{Harmark,Chern,Pappas} and references therein), the number of studies
regarding rotating BHs is very limited \cite{Meryem1,Meryem2,Ida,SChen,Jorge}%
. Even there have been very few studies devoted to the NAF rotating BHs \cite%
{Doukas,Meryem3,Li13}. This scarcity comes from the technical difficulty of
getting exact analytical solution (EAS) to the considered wave equation.
In fact, EAS method (see for example \cite{Harmark,Fern,Ngamp}) applies to
BH geometries which depend on a radial coordinate.

In this paper, we study GF, ACS, and DR of the RLDBH in 4-dimensions,
which is a solution to EMDA theory \cite{Clement}. To this end, we consider
the massless scalar particle and mainly follow the studies of \cite%
{Fern,Myung,Sakalli15} for using EAS method. It is worth noting that the
GF problem (without considering the problem of ACS and DR) of the RLDBH was firstly
considered (in broad strokes) in \cite{Li13}. However, as being stated in
the last paragraph of the conclusion of \cite{Li13}, the detailed analysis
of GF problem of RLDBH is not completed, and hence it deserves more
deeper research. Such an extension is one of the goals of the present paper.

The paper is organized as follows. Section 2 introduces RLDBH geometry,
and analyzes the Klein-Gordon equation (KGE) in this geometry. The angular
solution of the wave equation is given in Sect. 3. Section 4 is devoted to
the radial solution. GF, ACS, and DR computations are considered in Sect. 5.
The paper ends with a conclusion in Sect. 6.

\section{RLDBH in the EMDA Theory and KGE}

In the Boyer--Lindquist coordinates, the metric of RLDBH which is the
stationary axisymmetric EMDA BH \cite{Clement} is given by

\begin{equation}
ds^{2}=-fdt^{2}+\frac{dr^{2}}{f}+h\left[ d\theta ^{2}+\sin ^{2}\theta \left(
d\varphi -a\frac{dt}{h}\right) ^{2}\right] ,  \label{1}
\end{equation}

with the metric function

\begin{equation}
f=\frac{\Delta }{h},  \label{2}
\end{equation}

where $h=rr_{0}$ in which $r_{0}$ is a positive constant. In fact, $r_{0}$
is related to the background electric charge and finely
tunes the dilaton and axion \cite{Clement} fields, which are associated with
the dark matter \cite{AxDark,DilDark}. Besides

\begin{equation}
\Delta =(r-r_{+})(r-r_{-}),  \label{3}
\end{equation}

where $r_{+}$ and $r_{-}$ are the outer (event) and inner (Cauchy) horizons,
respectively, given by the zeros of $g_{tt}$:

\begin{equation}
r_{\pm }=M\pm \sqrt{M^{2}-a^{2}},  \label{4}
\end{equation}

where $M$ is associated with the quasilocal mass ($M_{QL}$) via $M=2M_{QL}$ 
\cite{BrYk}, and $a$ denotes the rotation parameter, which also tunes the
both dilaton and axion fields \cite{Clement}. One can immediately see from
Eq. (4) that having a BH is conditional on $M\geq a$. Otherwise, there is no
horizon and the spacetime corresponds to a BH with a naked singularity at $%
r=0$. The angular momentum ($J$) and $a$ are related in the following way: $%
ar_{0}=2J$. When $a$ vanishes, RLDBH reduces to its static form, the
so-called linear dilaton black hole (LDBH) metric. For studies of LDBH,
the reader is referred to \cite%
{Clement2,PSak,Sak1,Sak2,Sak22,SHP1,SHP2,Sak3,Sak4,HRL1,HRL2,EPL}. The area ($%
A_{BH}$), Hawking temperature ($T_{RLDBH}^{H}$) and angular velocity ($\Omega
_{H}$) at the horizon are found to be \cite{Sakalli15}

\begin{equation}
A_{BH}=4\pi r_{0}r_{+},  \label{5}
\end{equation}

\begin{equation}
T_{RLDBH}^{H}=\frac{\kappa }{2\pi }=\left. \frac{\partial _{r}f}{4\pi }%
\right\vert _{r=r_{+}}=\frac{r_{+}-r_{-}}{4\pi r_{0}r_{+}},  \label{6}
\end{equation}

\begin{equation}
\Omega _{H}=2\frac{J}{r_{0}^{2}r_{+}}=\frac{a}{r_{0}r_{+}}.  \label{7}
\end{equation}

It is worth noting that $T_{RLDBH}^{H}$ vanishes at the extremal limit $M=a$%
, i.e., $r_{+}=r_{-}$. Moreover, as $a\rightarrow 0$ ($r_{-}\rightarrow 0$), 
$T_{RLDBH}^{H}\rightarrow T_{LDBH}^{H}=\frac{1}{4\pi r_{0}}$ which is
independent of the mass of the BH, and points an isothermal HR \cite%
{Clement2,SHP1}.

The massless KGE equation in a curved spacetime is given by

\begin{equation}
\partial _{\mu }\left( \sqrt{-g}g^{\mu \upsilon }\partial _{\upsilon }\Psi
\right) =0.  \label{8}
\end{equation}

It is straightforward to show that Eq. (8) separates for the solution ansatz
of the form $\Psi =\psi (r,\theta )e^{i(m\varphi -\omega t)}$, where $m$ and 
$\omega $ are constant associated with rotation in the $\varphi $-direction
and frequency, respectively. Thus, we can obtain the following master
equation

\begin{equation}
\partial _{r}\left( \Delta \partial _{r}\psi \right) +\frac{\partial
_{\theta }(\sin \theta \partial _{\theta }\psi )}{\sin \theta }+\left[ \frac{%
\left( h\omega -am\right) ^{2}}{hf}-\left( \frac{m}{\sin \theta }\right) ^{2}%
\right] \psi =0.  \label{9}
\end{equation}

If we let $\psi =R(r)\Theta (\theta )$, Eq. (9) is separated into radial and
angular equations as follows

\begin{equation}
\Delta \partial _{r}\left( \Delta \partial _{r}R\right) +\left[ (h\omega
-am)^{2}-\lambda \Delta \right] R=0,  \label{10}
\end{equation}

\begin{equation}
\partial _{\theta }(\sin \theta \partial _{\theta }\Theta )+\sin \theta 
\left[ \lambda -\left( \frac{m}{\sin \theta }\right) ^{2}\right] \Theta =0,
\label{11}
\end{equation}

where $\lambda $ denotes the eigenvalue.

\section{Solution of the Angular Equation}

In order to have the general solution to Eq. (10), we introduce a new
dimensionless variable as follows

\begin{equation}
z=\frac{1-\cos \theta }{2},  \label{12}
\end{equation}

so that Eq. (11) becomes

\begin{equation}
z(1-z)\partial _{yy}\Theta +(1-2z)\partial _{y}\Theta +\left[ \frac{%
4z(z-1)\lambda +m^{2}}{4z(z-1)}\right] \Theta =0.  \label{13}
\end{equation}

One can rewrite the factor of $\Theta $ in the third term of Eq. (13) as
follows

\begin{equation}
\frac{4\lambda z(z-1)+m^{2}}{4z(z-1)}=\lambda -\frac{m^{2}}{4z}+\frac{m^{2}}{%
4(z-1)},  \label{14}
\end{equation}

Letting 
\begin{equation}
\Theta =\left( \frac{1-z}{z}\right) ^{\frac{m}{2}}\Phi (z),  \label{15}
\end{equation}

equation (13) is transformed into

\begin{equation}
z(1-z)\partial _{zz}\Phi +\left[ \overline{c}-(1+\overline{a}+\overline{b})z%
\right] \partial _{z}\Phi -\overline{a}\overline{b}\Phi =0.  \label{16}
\end{equation}

The above equation resembles the standard hypergeometric equation \cite{AS1}
whose solution is given by

\begin{equation}
\Phi =C_{1}F\left( \overline{a},\overline{b};\overline{c};z\right)
+C_{2}z^{1-\overline{c}}F\left( \overline{a}-\overline{c}+1,\overline{b}-%
\overline{c}+1;2-\overline{c};z\right) ,  \label{17}
\end{equation}

where $F\left( \overline{a},\overline{b};\overline{c};z\right) $ is the
standard (Gaussian) hypergeometric function \cite{AS1}, and $C_{1}$, $C_{2}$
are integration constants. By performing a few algebraic manipulations, one
obtains the following identities

\begin{equation}
\overline{a}=\frac{1}{2}\left( 1-\sqrt{4\lambda +1}\right) ,  \label{18}
\end{equation}

\begin{equation}
\overline{b}=1-\overline{a}=\frac{1}{2}\left( 1+\sqrt{4\lambda +1}\right) ,
\label{19}
\end{equation}

\begin{equation}
\overline{c}=1-m.  \label{20}
\end{equation}

Consequently, the general angular solution reads

\begin{equation}
\Theta =C_{1}\left( \frac{1-z}{z}\right) ^{\frac{m}{2}}F\left( \overline{a},%
\overline{b};\overline{c};z\right) +C_{2}\left[ z(1-z)\right] ^{\frac{m}{2}%
}F\left( \overline{a}-\overline{c}+1,\overline{b}-\overline{c}+1;2-\overline{%
c};z\right) .  \label{21}
\end{equation}

However, we need the normalized angular solution \cite{Mitsuo}. For this purpose, we
initially set $C_{2}=0$, and assign the eigenvalue to

\begin{equation}
\lambda =l(l+1),  \label{22}
\end{equation}

where $l=0,1,2,3..$. Using the following transformation \cite{NIST1}

\begin{equation}
F(\overline{a},1-\overline{a};\overline{c};z)=\left( \frac{-z}{1-z}\right)
^{(1-\overline{c})/2}P\left( -\overline{a},1-\overline{c},1-2z\right) ,
\label{23}
\end{equation}

where $P$ denotes the associated Legendre polynomials \cite{AS1}, we re-express

\begin{equation}
\Theta =\widehat{C}_{1}P(l,m,1-2z),  \label{24}
\end{equation}

which can rewritten as

\begin{equation}
\Theta =\widehat{C}_{1}P(l,m,\cos \theta ),  \label{25}
\end{equation}

where $\widehat{C}_{1}=C_{1}(-1)^{\frac{m}{2}}$. Employing the orthonomality
relation \cite{Mitsuo} for the associated Legendre functions and taking $%
e^{im\varphi }$ into account, we obtain the physical angular solution in
terms of the spherical harmonics \cite{NIST1}:

\begin{equation}
Y_{l,m}(\theta ,\varphi )=e^{im\varphi }\sqrt{\frac{\left( 2l+1\right)
\left( l-m\right) !}{4\pi (l+m)!}}P(l,m,\cos \theta ),  \label{26}
\end{equation}

where the index $l$ corresponds to the well-known azimuthal quantum number,
and $m$ denotes the magnetic quantum number (integer) with $-l\leq m\leq l$.

\section{Solution of the Radial Equation}

In this section, we give the exact analytical solution of the radial
equation. While doing this, we borrow the ideas from the recent study of
Sakalli \cite{Sakalli15}, which is about the area quantization of the RLDBH.

Using the following substitution

\begin{equation}
y=\frac{r-r_{+}}{r_{-}-r_{+}},  \label{27}
\end{equation}

one can rewrite Eq. (10) as

\begin{equation}
y(1-y)\partial _{yy}R+(1-2y)\partial _{y}R+\left( \frac{A^{2}}{y}-\frac{B^{2}%
}{1-y}+C\right) R=0,  \label{28}
\end{equation}

by which the constants are given by%
\begin{equation}
A=\frac{\omega r_{0}r_{+}-ma}{r_{+}-r_{-}}=\frac{\widetilde{\omega }}{%
2\kappa },  \label{29}
\end{equation}

\begin{equation}
B=i\left( \frac{\omega r_{0}r_{-}-ma}{r_{+}-r_{-}}\right) ,  \label{30}
\end{equation}

\begin{equation}
C=\lambda -\omega ^{2}r_{0}^{2}.  \label{31}
\end{equation}

where $\widetilde{\omega }=\omega -m\Omega _{H}$ denotes the wave frequency
detected by the observer rotating with the horizon \cite{Clement,Star2}.
Using the following ansatz

\begin{equation}
R(y)=\left( -y\right) ^{iA}(1-y)^{-B}\mathcal{R}(y),  \label{32}
\end{equation}

equation (28) reduces to the hypergeometric differential equation \cite{AS1}

\begin{equation}
y(1-y)\partial _{yy}\mathcal{R}\left[ \widehat{c}-(1+\widehat{a}+\widehat{b}%
)y\right] \partial _{y}\mathcal{R}-\widehat{a}\widehat{b}\mathcal{R}=0,
\label{33}
\end{equation}

whose solution has the same form as Eq. (17). The parameters of Eq. (33) are
given by

\begin{equation}
\widehat{a}=\frac{1}{2}(1+\sqrt{1+4C})+iA-B,  \label{34}
\end{equation}

\begin{equation}
\widehat{b}=\frac{1}{2}(1-\sqrt{1+4C})+iA-B,  \label{35}
\end{equation}%
\begin{equation}
\widehat{c}=1+2iA,  \label{36}
\end{equation}

or

\begin{equation}
\widehat{a}=\frac{1}{2}+ir_{0}\left( \omega +\widehat{\omega }\right) ,
\label{37}
\end{equation}

\begin{equation}
\widehat{b}=\frac{1}{2}+ir_{0}\left( \omega -\widehat{\omega }\right) ,
\label{38}
\end{equation}

\begin{equation}
\widehat{c}=1+i\frac{\widetilde{\omega }}{\kappa },  \label{39}
\end{equation}

where

\begin{eqnarray}
\widehat{\omega } &=&\sqrt{\omega ^{2}-\frac{1}{r_{0}^{2}}\left( \lambda +%
\frac{1}{4}\right) },  \notag \\
&=&\sqrt{\omega ^{2}-\left( \frac{l+1/2}{r_{0}}\right) ^{2}}.  \label{40}
\end{eqnarray}

Inspired by the work of Fernando \cite{Fern}, we only consider the modes
having the real values of $\widehat{\omega }$ (analogous to the
Breitenlohner-Freedman bound \cite{BFbound}), which correspond to the
following frequencies

\begin{equation}
\omega \geq \frac{l+1/2}{r_{0}}.  \label{41}
\end{equation}

Finally, the general radial solution of Eq. (28) becomes

\begin{eqnarray}
R(y) &=&D_{1}\left( -y\right) ^{iA}(1-y)^{-B}F\left( \widehat{a},\widehat{b};%
\widehat{c};y\right) +  \notag \\
&&D_{2}\left( -y\right) ^{-iA}(1-y)^{-B}F\left( \widehat{a}-\widehat{c}+1,%
\widehat{b}-\widehat{c}+1;2-\widehat{c};y\right) ,  \label{42}
\end{eqnarray}

where $D_{1}$, $D_{2}$ are integration constants.

\section{GF, ACS, and DR Computations}

In general, GF depends on the behavior of the radial function both at
the horizon and at asymptotic infinity. Namely, GF is defined by the
flux $\mathcal{F}$, and ACS and DR are the consequences of GF.

Using Taylor series expansion, we can expand the metric function $f$ around
the event horizon ($r\rightarrow r_{+}$ or $y\rightarrow 0$):

\begin{align}
f_{NH}& \simeq \left. \partial _{r}f\right\vert
_{r=r_{+}}(r-r_{+})+O(r-r_{+})^{2},  \notag \\
& \simeq 2\kappa \left( r_{+}-r_{-}\right) x,  \label{43}
\end{align}

where $x=-y$, and subscript NH denotes the near-horizon. Using the
definition of the tortoise coordinate ($r^{\ast }$) \cite{MTBH1}, we obtain
NH form of $r^{\ast }$ as

\begin{equation}
r^{\ast }\simeq \int \frac{dr}{f_{NH}}=\int \frac{dx}{2\kappa x}=\frac{1}{%
2\kappa }\ln x,  \label{44}
\end{equation}

which yields%
\begin{equation}
x=e^{2\kappa r_{\ast }}.  \label{45}
\end{equation}

Thus, we obtain the following NH solution

\begin{equation}
R_{NH}\simeq D_{1}e^{i\widetilde{\omega }r_{\ast }}+D_{2}e^{-i\widetilde{%
\omega }r_{\ast }}.  \label{46}
\end{equation}

Correspondingly, NH partial wave is given by

\begin{equation}
\Psi _{NH}\simeq D_{1}e^{i\left( \widetilde{\omega }r_{\ast }-\omega
t\right) }+D_{2}e^{-i\left( \widetilde{\omega }r_{\ast }+\omega t\right) }.
\label{47}
\end{equation}

Now, we interpret $D_{1}$ and $D_{2}$ as the amplitudes of the NH outgoing
and ingoing waves, respectively. However, we need the physical solution that
reduces to the ingoing wave at the horizon. Therefore, we simply set $%
D_{1}=0 $, and the general solution (42) reduces to

\begin{equation}
R(y)=D_{2}\left( x\right) ^{-iA}(1+x)^{-B}F\left( \widehat{a}-\widehat{c}+1,%
\widehat{b}-\widehat{c}+1;2-\widehat{c};y\right) .  \label{48}
\end{equation}

Near SI $(r\rightarrow \infty )$, the metric function $f$ becomes

\begin{equation}
f_{SI}\simeq \frac{r}{r_{0}},  \label{49}
\end{equation}

and the tortoise coordinate reads%
\begin{equation}
r^{\ast }\simeq \int \frac{dr}{f_{SI}}=r_{0}\ln r.  \label{50}
\end{equation}

Therefore, the variable $x$ at SI behaves as

\begin{equation}
x\simeq \frac{r}{r_{+}-r_{-}}=\frac{e^{\frac{r^{\ast }}{r_{0}}}}{r_{+}-r_{-}}%
.  \label{51}
\end{equation}

Thus at SI while $r$,$r^{\ast }\rightarrow \infty $, $x\rightarrow \infty $
(or $y\rightarrow -\infty $). Using the following inverse transformation
formula \cite{AS1,NIST1}

\begin{eqnarray}
F(\widetilde{a},\widetilde{b};\widetilde{c};z) &=&(-z)^{-\widetilde{a}}\frac{%
\Gamma (\widetilde{c})\Gamma (\widetilde{b}-\widetilde{a})}{\Gamma (%
\widetilde{b})\Gamma (\widetilde{c}-\widetilde{a})}F(\widetilde{a},%
\widetilde{a}+1-\widetilde{c};\widetilde{a}+1-\widetilde{b};1/z)+  \notag \\
&&(-z)^{-\widetilde{b}}\frac{\Gamma (\widetilde{c})\Gamma (\widetilde{a}-%
\widetilde{b})}{\Gamma (\widetilde{a})\Gamma (\widetilde{c}-\widetilde{b})}F(%
\widetilde{b},\widetilde{b}+1-\widetilde{c};\widetilde{b}+1-\widetilde{a}%
;1/z),  \label{52}
\end{eqnarray}

and substituting Eq. (51) in Eq (48), after performing a few algebraic
manipulations, we obtain the partial wave at SI as follows

\begin{equation}
\Psi _{SI}\simeq \frac{\sqrt{r_{+}-r_{-}}}{\sqrt{r}}\left[ E_{1}\left(
r_{+}-r_{-}\right) ^{ir_{0}\widehat{\omega }}e^{-i(r^{\ast }\widehat{\omega }%
+\omega t)}\right. +\left. E_{2}\left( r_{+}-r_{-}\right) ^{-ir_{0}\widehat{%
\omega }}e^{i(r^{\ast }\widehat{\omega }-\omega t)}\right] ,  \label{53n}
\end{equation}

where $E_{1}$ and $E_{2}$ correspond to the amplitudes of the asymptotic
ingoing and outgoing waves, respectively. The relations between ($%
E_{1},E_{2} $) and $D_{2}$ are

\begin{equation}
E_{1}=\frac{\Gamma (2-\widehat{c})\Gamma (\widehat{b}-\widehat{a})}{\Gamma (%
\widehat{b}-\widehat{c}+1)\Gamma (1-\widehat{a})}D_{2},  \label{54n}
\end{equation}

\begin{equation}
E_{2}=\frac{\Gamma (2-\widehat{c})\Gamma (\widehat{a}-\widehat{b})}{\Gamma (%
\widehat{a}-\widehat{c}+1)\Gamma (1-\widehat{b})}D_{2}.  \label{55n}
\end{equation}

The current ($J^{C}$) \cite{Ding} is defined as

\begin{equation}
J^{C}=\frac{1}{2i}\left( \overline{\Psi }\partial _{r^{\ast }}\Psi -\Psi
\partial _{r^{\ast }}\overline{\Psi }\right) ,  \label{56}
\end{equation}

where the bar over a quantity denotes the complex conjugation. The current
is the flux per unit coordinate area \cite{Ding}. Therefore, the ingoing
flux at the horizon is

\begin{equation}
\mathcal{F}_{NH}=\frac{A_{BH}}{2i}\left( \overline{\Psi }_{NH}\partial
_{r^{\ast }}\Psi _{NH}-\Psi _{NH}\partial _{r^{\ast }}\overline{\Psi }%
_{NH}\right) =-4\pi \widetilde{\omega }r_{+}r_{0}\left\vert D_{2}\right\vert
^{2}.  \label{57}
\end{equation}

Similarly, the incoming flux at SI is given by

\begin{equation}
\mathcal{F}_{SI}=\frac{A_{BH}}{2i}\left( \overline{\Psi }_{SI}\partial
_{r^{\ast }}\Psi _{SI}-\Psi _{SI}\partial _{r^{\ast }}\overline{\Psi }%
_{SI}\right) =-4\pi \widehat{\omega }\left( r_{+}-r_{-}\right)
r_{0}\left\vert D_{2}\right\vert ^{2}\left\vert \tau \right\vert ^{2},
\label{58}
\end{equation}

where

\begin{equation}
\tau =\frac{\Gamma (2-\widehat{c})\Gamma (\widehat{b}-\widehat{a})}{\Gamma (%
\widehat{b}-\widehat{c}+1)\Gamma (1-\widehat{a})}.  \label{59}
\end{equation}

The GF (or the absorption probability) is given by \cite{Harmark,Fern,Campu}

\begin{equation}
\gamma =\frac{\mathcal{F}_{NH}}{\mathcal{F}_{SI}}=\frac{\widetilde{\omega }%
r_{+}}{\widehat{\omega }\left( r_{+}-r_{-}\right) }\left\vert \tau
\right\vert ^{-2}.  \label{60}
\end{equation}

By substituting Eqs. (37-39) into the above equation, and using the
following identities for the Gamma functions \cite{AS1}

\begin{equation}
\left\vert \Gamma (iv)\right\vert ^{2}=\frac{\pi }{v\sinh \left( \pi
v\right) },  \label{61}
\end{equation}

\begin{equation}
\left\vert \Gamma (1+iv)\right\vert ^{2}=\frac{\pi v}{\sinh \left( \pi
v\right) },  \label{62}
\end{equation}

\begin{equation}
\left\vert \Gamma (\frac{1}{2}+iv)\right\vert ^{2}=\frac{\pi }{\cosh \left(
\pi v\right) },  \label{63}
\end{equation}

equation (60) becomes

\begin{equation}
\gamma =\frac{\sinh \left( \frac{\pi \widetilde{\omega }}{\kappa }\right)
\sinh \left( 2\pi r_{0}\widehat{\omega }\right) }{\cosh \left( \frac{\pi 
\widetilde{\omega }}{\kappa }-\pi \alpha \right) \cosh \left( \pi \beta
\right) },  \label{64}
\end{equation}

where

\begin{equation}
\alpha =r_{0}(\omega -\widehat{\omega }),  \label{65}
\end{equation}

\begin{equation}
\beta =r_{0}(\omega +\widehat{\omega }).  \label{66}
\end{equation}

One can also show that the above expression recasts in

\begin{equation}
\gamma =\frac{\left( e^{\frac{2\pi \widetilde{\omega }}{\kappa }}-1\right)
\left( e^{4\pi r_{0}\widehat{\omega }}-1\right) }{\left[ e^{2\pi \left( 
\frac{\widetilde{\omega }}{\kappa }-\alpha \right) }+1\right] \left( e^{2\pi
\beta }+1\right) }.  \label{67}
\end{equation}

The partial ACS for a 4-dimensional BH is given by \cite{Unruh,Chin}

\begin{equation}
\sigma _{abs}^{l,m}=\frac{\left( 2l+1\right) \pi }{\omega ^{2}}\gamma ,
\label{68}
\end{equation}

which is equal to

\begin{equation}
\sigma _{abs}^{l,m}=\frac{\left( 2l+1\right) \pi \left( e^{\frac{2\pi 
\widetilde{\omega }}{\kappa }}-1\right) \left( e^{4\pi r_{0}\widehat{\omega }%
}-1\right) }{\omega ^{2}\left[ e^{2\pi \left( \frac{\widetilde{\omega }}{%
\kappa }-\alpha \right) }+1\right] \left( e^{2\pi \beta }+1\right) }.
\label{69}
\end{equation}

Meanwhile, one can also compute the total ACS \cite{Unruh,Chin} via the
following expression

\begin{equation}
\sigma _{abs}^{Total}=\sum\limits_{l=0}^{\infty }\sigma _{abs}^{l,m}.
\label{70}
\end{equation}

It is worth noting that for a $S$-wave ($l=0$, $\therefore $ $m=0$) $\sigma
_{abs}^{0,0}$ was shown to be equal to the area of the BH while $\omega
\rightarrow 0$ by Das et al \cite{Das}. However, in this study (as in the
case of 3-dimensional static charged dilaton BH \cite{Fern,Myung}) condition (41)
restricts $\omega $ to \emph{positive numbers}. So, we can not take $\omega
\rightarrow 0$\ limit to prove the result of \cite{Das}. Moreover, $\omega
\rightarrow 0$ limit makes $\widehat{\omega }$ imaginary in Eq. (40). Hence,
the incoming flux (58) at SI vanishes \cite{Fern,Myung}. This problem could be overcome by the
procedure described in \cite{Bir}, which computes $\sigma _{abs}^{0,0}$ at
the low frequency.

$\sigma _{abs}^{0,0}$, $\sigma _{abs}^{1,1}$, $\sigma _{abs}^{2,0}$ and $%
\sigma _{abs}^{3,-1}$ graphs are plotted in Fig. (1). The graphs are drawn
for $\omega \geq \frac{l+1/2}{r_{0}}$. While the locations of the peaks (local maximums) on
the $\omega $-axis (which are very close to starting frequency $\omega _{l}=%
\frac{l+1/2}{r_{0}}$) of $\sigma _{abs}^{l,m}$ shift towards right with
increasing $l$-value, however the peak values decrease when $l$ gets bigger values.
Besides, every $\sigma _{abs}^{l,m}$ engrossingly follows the same line to
die down while $\omega \rightarrow \infty $ (as such in \cite{Myung}).
Although it is not illustrated here, however it is deduced from our detailed analyzes that the behaviors of $\sigma _{abs}^{l,m}$ are almost same for the other configurations (for
different values of $M$, $a$, and $r_{0}$) of the RLDBH. As a final remark,
we have not observed any negative $\sigma _{abs}^{l,m}$ behavior in our
analyzes. So, the superradiance \cite{SRad} phenomenon ( showing up itself
for the waves with $\omega <m\Omega _{H}$) does not occur. This can be best
seen by imposing the superradiance condition:

\begin{equation}
\omega _{l}<m\Omega _{H}\text{ \ \ \ }\rightarrow \text{ \ \ \ }l+1/2<\frac{%
ma}{M+\sqrt{M^{2}-a^{2}}}.  \label{71}
\end{equation}

However, the above inequality holds whenever $m>l,$ which is an inadmissible
case. Namely, the starting frequency $\omega _{l}$ is always greater than $%
m\Omega _{H}.$
\begin{figure}[H]
\centering
\includegraphics[scale=.75]{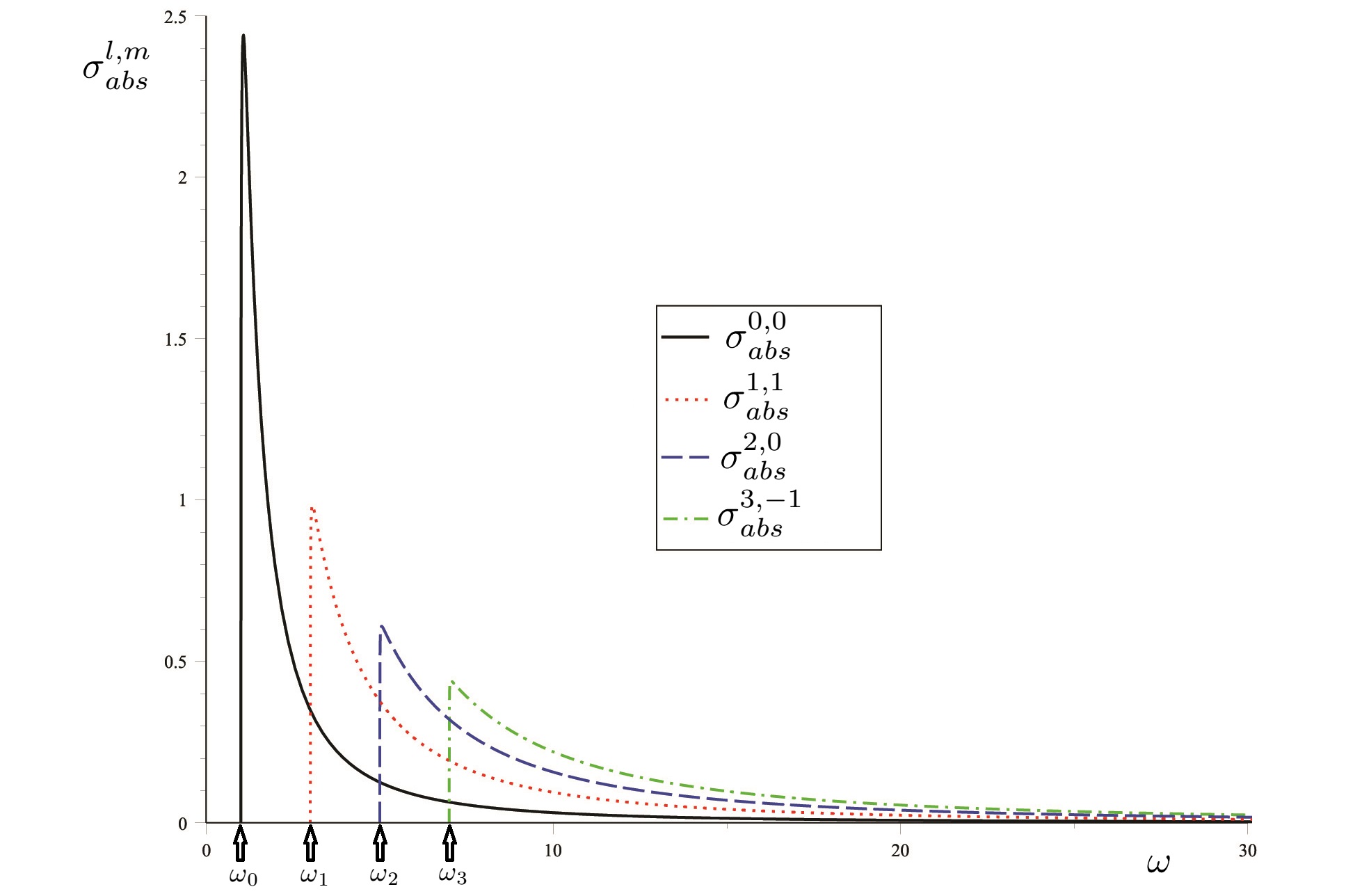}
\caption{Plots of the partial ACS $(\sigma _{abs}^{l,m})$ versus frequency $%
\omega $. The plots are governed by Eq. (69). The configuration of the RLDBH
is as follows $M=1$, $a=0.2$, and $r_{0}=0.5$. The starting frequencies are
determined by $\omega _{l}=\frac{l+1/2}{r_{0}}$.}
\end{figure}

On the other hand, DR for a rotating BH is given by \cite{Chin}

\begin{equation}
\Gamma ^{l,m}=\frac{\sigma _{abs}^{l,m}}{e^{\widetilde{\omega }/T_{H}}-1}.
\label{72}
\end{equation}

Hence, using Eq. (69), we obtain

\begin{equation}
\Gamma ^{l,m}=\frac{\pi \left( e^{4\pi r_{0}\widehat{\omega }}-1\right) }{%
\omega ^{2}\left[ e^{2\pi \left( \frac{\widetilde{\omega }}{\kappa }-\alpha
\right) }+1\right] \left( e^{2\pi \beta }+1\right) }.  \label{73}
\end{equation}

The four graphs for $\Gamma ^{0,0}$, $\Gamma ^{1,1}$, $\Gamma ^{2,0}$, and $%
\Gamma ^{3,-1}$ (each one with its own scale factor: $\times 10^{-n}$) are
depicted in Fig. (2). It is clear that similar to $\sigma _{abs}^{l,m}$ the
peaks of $\Gamma ^{l,m}$ become smaller for increasing $l$-values. Also DRs quickly deplete themselves with increasing $\omega $-values (similar plots can
be seen in \cite{Fern}).
\begin{figure}[H]
\centering
\includegraphics[scale=.75]{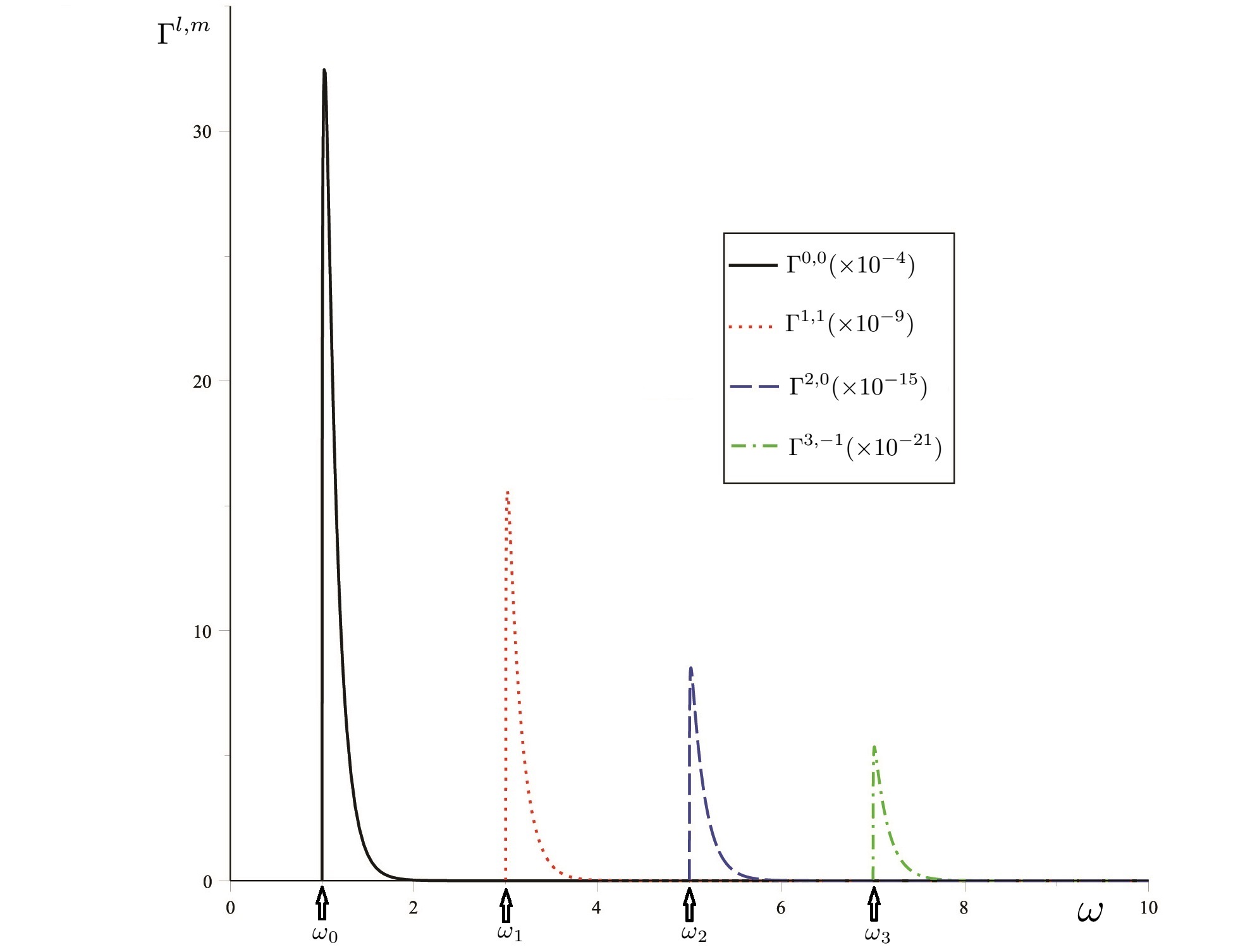}
\caption{Plots of the DR $(\Gamma ^{l,m})$ versus frequency $\omega $. The
plots are governed by Eq. (73). The configuration of the RLDBH is as follows 
$M=1$, $a=0.2$, and $r_{0}=0.5$. The starting frequencies are
determined by $\omega _{l}=\frac{l+1/2}{r_{0}}$.}
\end{figure}

\section{Conclusion}

The propagation and dynamic evolution of the massless scalar field in the
background of the RLDBH with the condition of Eq. (41) (inspired by the
study of \cite{Fern}) have been studied. We have obtained an exact solution
for the KGE in the whole geometry. Then, imposing the boundary conditions at
the event horizon and SI, we have found an analytical expression (67) for
the GF of the scalar fields propagating in the RLDBH spacetime. In the
sequel, we have computed the exact ACS and DR for the RLDBHs. In Figs. (1)
and (2), we have demonstrated that both ACS and DR (starting from $\omega
_{l}=\frac{l+1/2}{r_{0}}$) expeditiously reach to their peak values and then
approach to zero with increasing frequency. However, the peaks of DRs
are more smaller comparing with the ACS ones for the same configured RLDBH.
Moreover, it is seen that negative ACS, which implies the superradiance \cite%
{SRad} is not possible for the waves satisfying the condition of Eq. (41).\
This is due to the inequality given in Eq. (71). Moreover, for the
lower frequencies ($0\leq \omega <\frac{l+1/2}{r_{0}}$) we have shown that
our calculations lead to zero incoming flux (58) at SI (similar to \cite%
{Fern,Myung}). To overcome this problem, one could follow the steps given in \cite%
{Bir}. However, this aspect was beyond the scope of the present paper.

In future, we plan to extend our analysis to the fermion perturbations,
which are performed by the virtue of the Dirac equation in 4-dimensions \cite%
{MTBH1,JMP}. In this way, we will analyze the ACS and DR of the RLDBHs using
fermion GFs.

\section*{Acknowledgement}
The authors are grateful to the editor and anonymous referee for their
valuable comments and suggestions to improve the paper.

\end{document}